\documentclass[useAMS,usenatbib,fleqn]{mn2e}
\usepackage{epsfig}
\usepackage{color} 
\usepackage{natbib}
\usepackage[colorlinks=true,citecolor=blue]{hyperref}
\def\rrm{rad m$^{-2}$}

\def\ewr{$EW_{r}$}
\def\mgi{Mg~{\sc i}} 
\def\mgii{Mg~{\sc ii}~} 
\def\mgiia{Mg~{\sc ii}$\lambda$2796} 
\def\mgiib{Mg~{\sc ii}$\lambda$2803} 

\def\feii{Fe~{\sc ii}~} 
 
\def\feiia{Fe~{\sc ii}$\lambda$2600~} 

\def\civ{C~{\sc iv}~}

\def\zem{$z_{\rm em}$~} 
 
\def\lya{Ly$\alpha$}

\def\chisq{$\chi^{2}$} 
\def\kms{km.s$^{-1}$} 
\title[RRM and intervening \mgii absorbers]{
Dependence of Residual Rotation Measure (RRM) on Intervening
\mgii Absorbers  at Cosmic Distances}
\author[$Joshi$ \& $Chand$ ]{Ravi Joshi$^{1}$\thanks{E-mail: ravi@aries.res.in(RJ);
hum@aries.res.in(HC)}, Hum Chand$^{1}$$^{\star}$ \\ $^{1}$Aryabhatta Research Institute of observational sciencES (ARIES),
Manora Peak, Nainital $-$ 263129, India\\\\}

\begin{document}
\date{Accepted ---. Received ---; in original form ---}

\pagerange{\pageref{firstpage}--\pageref{lastpage}} \pubyear{2013}

\maketitle

\label{firstpage}
\begin{abstract}

 We investigate the dependence of residual rotation measure (RRM) on
 intervening absorption systems at cosmic distances by using a large
 sample of 539 SDSS quasars in conjunction with the available rotation
 measure catalog at around 21cm wavelength. We found an excess
 extragalactic contribution in standard deviation of observed RRM
 ($\sigma_{rrm}$) of about 8.11$\pm4.83$~\rrm\ in our sample with
 intervening \mgii absorber as compare to the sample without \mgii
 absorber. Our results suggest that intervening
 absorbers could contribute to the enhancement of RRM at around 21cm
 wavelength, as was found earlier for RM measurements at around 6cm
 wavelength.
\end{abstract}


\section{Introduction}
\label{sec:intro_frm}

Magnetic field plays a key role in the structural and dynamical
evolution of the Universe
\citep[e.g.,][]{Mestel1984A&A...136...98M,Rees1987QJRAS..28..197R},
but there are no methods for its direct measurement. Faraday Rotation
(FR) is one of the powerful probes to measure the strength of magnetic
field over the cosmic time scale
~\citep[e.g,][]{Bernet2008Natur.454..302B, Bernet2012ApJ...761..144B,
  Hammond2012arXiv1209.1438H, You2003AcASn..44S.155Y,
  Welter1984ApJ...279...19W,
  Kronberg2008ApJ...676...70K,Kronberg1982ApJ...263..518K,Kronberg1977A&A....61..771K,
  Kronberg1976Natur.263..653K}. This Rotation Measure (RM) is defined
 as the change in observed polarization angle ($\Delta \chi_{0}$)
per unit change in observed wavelength square ($\Delta
\lambda_{0}^{2}$). For a linearly polarized radio source at
cosmological redshift ($z_{s}$) it is given
by~\citep{Bernet2012ApJ...761..144B}:

\begin{equation}
\label{eq:defRM}
RM(z_{s}) = \frac{{\Delta \chi_0 }}{{\Delta \lambda _0^{2} }} = 8.1\times
10^{5} \int\limits_{z_{s}}^{0} \frac{n_e
(z)B_\parallel(z)}{(1+z)^{2}}\frac{dl}{dz}dz,
\end{equation}

where RM is in units of~\rrm, the free electron number density, $n_e$,
is in $cm^{-3}$, the magnetic field component along the line-of-sight,
$B_{||}$, is in Gauss, and the comoving path increment per unit
redshift, $dl/dz$, is in parsec. However, the observed  RM has
contributions from two components, namely extragalactic radio
  source and ionized medium of our Galaxy. As a result of this, it is
not straight forward to quantify the FR contribution coming from the
extragalactic radio source. This extragalactic component also includes
the contributions from intervening galaxies and/or their halos,
  protogalaxies, intergalactic clouds, an intracluster gas consisting
  of widespread coexpanding diffuse intergalactic medium and intrinsic
  to the quasar. Therefore, the extragalactic RM can only be studied
in terms of the residual rotation measure (RRM), after removing the
Galactic component from the observed RM. \par

Earlier studies of RM on its redshift evolution has shown that the RM
dispersion of quasars increases at high
redshift~\citep[][]{Kronberg2008ApJ...676...70K,
  Kronberg1976Natur.263..653K, Rees1972A&A....19..189R}, in contrast
to the $(1+z)^{-2}$ dilution effect on RM (e.g., see
Eq.~\ref{eq:defRM}). This led to the conclusion that the magnetic
field strength as traced by the RM of high redshift galaxies is at
least comparable to the current
epoch~\citep{Kronberg2008ApJ...676...70K}. The possible origin of this
high magnetic field at cosmic distances remain ambiguous, however,
such high fields could be either intrinsic to the quasars (e.g.,
arising in its immediate environment) or due to the intervening
environments along the lines-of-sight between the polarized source and
the observer.~\citet{Bernet2008Natur.454..302B} have probed the latter
possibility from the analysis of  high resolution optical
spectra of 76 quasars. They have shown that the quasars with strong
\mgii absorption line systems are unambiguously associated with larger
RM, inferred from their RM observation  at around 6cm wavelength.
In other words, the major contribution to the extragalactic component
of observed RM comes from the intervening galaxies. However, in
contrast to the case with \mgii absorption systems having rest frame
equivalent width (\ewr)  greater than
0.3\AA,~\citet{Bernet2010ApJ...711..380B} have shown that the weaker
systems do not contribute significantly to the observed RM of the
background quasars. The above discrepancy is attributed to the higher
impact parameters of weak systems compared to strong ones. \par

Recently,~\citet{Hammond2012arXiv1209.1438H} made a catalog of RRM
available for 3651 radio sources that  were observed at around
  21cm. They reported an observed standard deviation in RRM of 23.2
\rrm\  from a mixed sample of quasars and galaxies having a
mixture of sightlines with and without \mgii absorber.  Further,
  subtracting the possible errors contributing to this measured
standard deviation, such as: (i) the measurement errors of individual
RM of 11 \rrm\ as given in the catalog
by~\citet{Taylor2009ApJ...702.1230T}; (ii) error associated with the
galactic rotation measure (GRM) calculations of 6
\rrm\ \citep{Oppermann2012A&A...542A..93O}; and (iii) 12$-$17
\rrm\ from the RM fluctuations on smaller angular scales than are
being sampled by above GRM~\citep{Stil2011ApJ...726....4S}. The
remaining contribution from extragalactic component is found to be
typically around 10$-$15 \rrm, similar to
~\citet[][]{Schnitzeler2010MNRAS.409L..99S}. Importantly, with this
extensive study, they could not reproduce any significant redshift
evolution of RRM as seen by  other
  studies~\citep[e.g.,][]{Kronberg2008ApJ...676...70K,
  Welter1984ApJ...279...19W}.
  
To explain the above discrepancy,~\citet{Bernet2012ApJ...761..144B}
have proposed a model consisting of partially inhomogeneous rotation
measure screen, which causes wavelength dependent depolarization. As a
result the depolarization in their model toward longer wavelength such
as close to 21cm used in~\citet{Hammond2012arXiv1209.1438H} data set
will be larger than the shorter wavelength  close to 6cm used
in~\citet{Bernet2008Natur.454..302B}, which has been attributed to the
above discrepancy. This was supported by their result that the RM
distribution with and without \mgii absorber do differ from data set
based on 6cm, unlike no such difference seen on their 21cm data set.
However, it should be noted that this data set having RM measurements
at both the above wavelengths (i.e., 21cm and 6cm) consist of only 54
radio source sightlines. In view of the important consequences of the
above results, it is very important to carry out the analysis by using
larger  sample size to detect and quantify any effect of
intervening absorbers on the extra-galactic component of RRM. This
forms the main motivation of our present work, by carrying out the
analyses for subsample of sightlines with and without \mgii absorber
in the parent sample of 567 SDSS quasars and by using the RRM-redshift
catalog provided by~\citet{Hammond2012arXiv1209.1438H}.

This paper is organized as follows. Section 2 describe the selection
of the samples of RRM data set, while Section 3 gives 
  methodology used in the analysis. In
Section 4, we present the results of our analysis, followed by
  the discussion and conclusions in Section 5.

\section{Sample}
\label{sec:sample_mgiidndz}

Our sample is collected from catalog produced by
~\citet{Hammond2012arXiv1209.1438H}, which consists of 3651
sources at high galactic latitude, having $|b| > 20^{\circ}$. The
catalog is constructed by assigning redshift for polarized radio
source cataloged by~\citet{Taylor2009ApJ...702.1230T} using an optical
database of, e.g., NED\footnote{NED=NASA/IPAC Extragalactic Database
  http://ned.ipac.caltech.edu} and SIMBAD\footnote{SIMBAD=Set of
  Identifications, Measurements and Bibliography for Astronomical
  Data; http://simbad.u-strasbg.fr/simbad/} and optical surveys namely
SDSS-DR8{\footnote{SDSS=Sloan Digital Sky survey DR8;
    http://skyserver/sdss3.org/dr8/en/}},
6DFGS{\footnote{6DFGS=Six-degree Field Galaxy Survey;
    http://www.-wfau.roe.ac.uk/6dfGS/}},
2dfGRS{\footnote{2dfGRS=Two-degree Field Galaxy Redshift Survey;
    http://www2.aao.gov.au/2dfGRS/}} and 2QZ/6QZ{\footnote{2QZ/6QZ=2df
    QSO Redshift Survey (2QZ) and 6df QSO Redshift Survey (6QZ);
    http://www.2dfquasar.org/Spec{$\_$}Cat/catalogue.html}} (e.g., see
their online Table 1). We have applied following criteria to select
our sample from the above mentioned catalog.

\begin{enumerate}
\item Firstly, we have restricted ourselves only to those radio
  sources whose optical association are assigned as quasar using SDSS
  database due to the advantage of  the availability of their
    spectra from SDSS archive. Of the 3651 source in the catalog
  of~\citet{Hammond2012arXiv1209.1438H}, this selection filter left us
  with 860 polarized radio sources having SDSS spectra. Out of them,
  we only consider the most common designated radio to optical
  association namely class `A', which represents an unresolved
  NVSS{\footnote{NVSS=NRAO Very Large Array Sky Survey;
      http://www.cv.nrao.edu/nvss/}} radio source (observed at 21cm)
  that align closely with the corresponding optical counterpart (i.e.,
  the NVSS-optical  offset $<$ 15 arcsec) and also have a
  FIRST{\footnote{FIRST=The VLA Faint Images of the Radio Sky at
      Twenty centimeters survey; http://sundog.stsci.edu/}}
  counterpart, which is used to make the radio-optical association.
  Further, the class `A' in~\citet{Hammond2012arXiv1209.1438H} is
  subdivided into seven subclasses A(i)-A(vii), depending on the
  radio morphology in FIRST survey. To minimize the uncertainty of
  radio and optical association, we have included only those
  subclasses having at the most 3 matches in FIRST image within 30
  arcsec of the optical position, leading to the exclusion of sources
  belonging to A(vi), which have more than 3 such associations. In
  addition,  we have included the class `B' sources, representing
  a similar situation to class `A' by using only NVSS, either due to
  the non-detection or the absence of data set in FIRST survey. This
  left us with a sample of 730 polarized radio sources having SDSS
  spectra.

\item Secondly, we have excluded all the quasars having emission
  redshift outside the range of $0.38 \le$ \zem $\le 2.3$. Here, the
  lower limit constrain on the quasar emission redshift is to ensure
  that their SDSS spectra having lowest wavelength about $3800$\AA,
  which allows to detect at least one \mgii doublet, if
  present. Similarly, upper \zem constrain is to ensure that \mgii
  emission line does not fall above the highest SDSS wavelength, which
  is $9200$\AA, so that ambiguity of any \mgii absorber falling above
  the spectral coverage can be avoided. These filters have reduced the
  sample to 615 quasars.

\item Finally, to minimize the uncertainty in the radio to
  optical association, we have removed those quasars
   for which the optical and radio sightlines are separated by more than
  7 arcsec (similar to \citealt{Bernet2008Natur.454..302B}), resulting in
  the sample of 567 quasars, for our analysis.

\end{enumerate}

\section{Analysis}
\label{sec:analysis_mgiidndz}

\subsection{Identification of \mgii absorption systems}
\label{subsec:mgii_iden_mgiidndz}

The identification of the \mgii absorption doublet in the normalized
continuum spectrum, was carried out using the procedure discussed in
detail by \citet[][under review]{2013MNRAStmp}. Briefly, the procedure
initially fits a continuum to the SDSS spectroscopic data, employing a
first principal component analysis (PCA) as a guess for the \lya\ and
\civ emission lines (i.e., from 1000\AA -2000\AA\ in the rest-frame).
Next, a b-spline algorithm was used to fit the underlying residual
continuum, which results roughly in a power-law with broad emission
features superposed. \par

The procedure automatically also searches for absorption features in
the normalized spectrum redward of \lya. The search was carried out
for absorption features, fitted with Gaussian profile by taking an
initial FWHM of 2.5 pixels, with the additional requirement that the
minimum separation between lines of doublet should be about 2 times
the FWHM. Out of all such cases, the final selection was made by
accepting only the lines which are above 3 times the rms($\sigma$)
noise in the spectrum. \par

The absorption features thus identified for each quasar, were searched
for absorption line pairs. For this purpose the procedure first
computed redshift of a given absorption feature, assuming it to be
\mgiia. The corresponding positions of the expected \mgiib\ and \feiia
lines were then inspected. The criterion for accepting a feature as
genuine \mgii absorption system, was that at least two of these three
lines must be present at the expected locations above a 3$\sigma$
threshold. Equivalent widths of all the accepted \mgii absorption
lines were then measured by summing the difference from unity of the
flux in the normalized observed-frame spectrum, within about 10 pixels
wide boxes ($\sim$11.51\AA\ in the observed frame) placed at the
centroids of the two \mgii lines. \par

As a further check, we also carried out a visual confirmation of each
absorption system identified via the above automated procedure. This
step is important since (i) our automated search do not carry out the
line profile matching and hence can result in over counting the \mgii
doublet rather than missing out any genuine system, and (ii) any
visually noticed uncertainty in the continuum level could have
significantly distorted the estimate of \ewr, rendering the
strong/weak classification of absorption systems unreliable. In this
process of visual scrutiny, we first looked for the strongest five
\feii lines corresponding to the candidate \mgii doublet. We then made
a velocity plot of \mgi, \mgii and the five \feii lines, so as to
match by eye the line profile and strength to that expected on the
basis of the line oscillator strengths. Thus, in the spectra of the
567 quasars, we visually inspected all 673 \mgii absorption systems
candidates and confirmed 256 of them. Continuum fitting of each
confirmed \mgii absorption system was then further checked by plotting
the fitted continuum to the spectral segment containing the \mgii
doublet. In all cases where the continuum fitting over the relevant
spectral segment seemed unsatisfactory, we refined the local continuum
fit and recomputed the \ewr({Mg~{\sc ii}}). To test the `intervening
hypothesis', we have removed those 28 quasars having associated \mgii
absorbers with relative velocity $<$ 5000~\kms\ in the spectra. This
leads to a final sample of 539 quasars, which is used in rest of our
analysis. Among these 539 quasars, 388 are without \mgii absorber,
while 119 have one and 32 have more than one \mgii absorber in their
spectra. \par

 \begin{figure*}
 \epsfig{figure=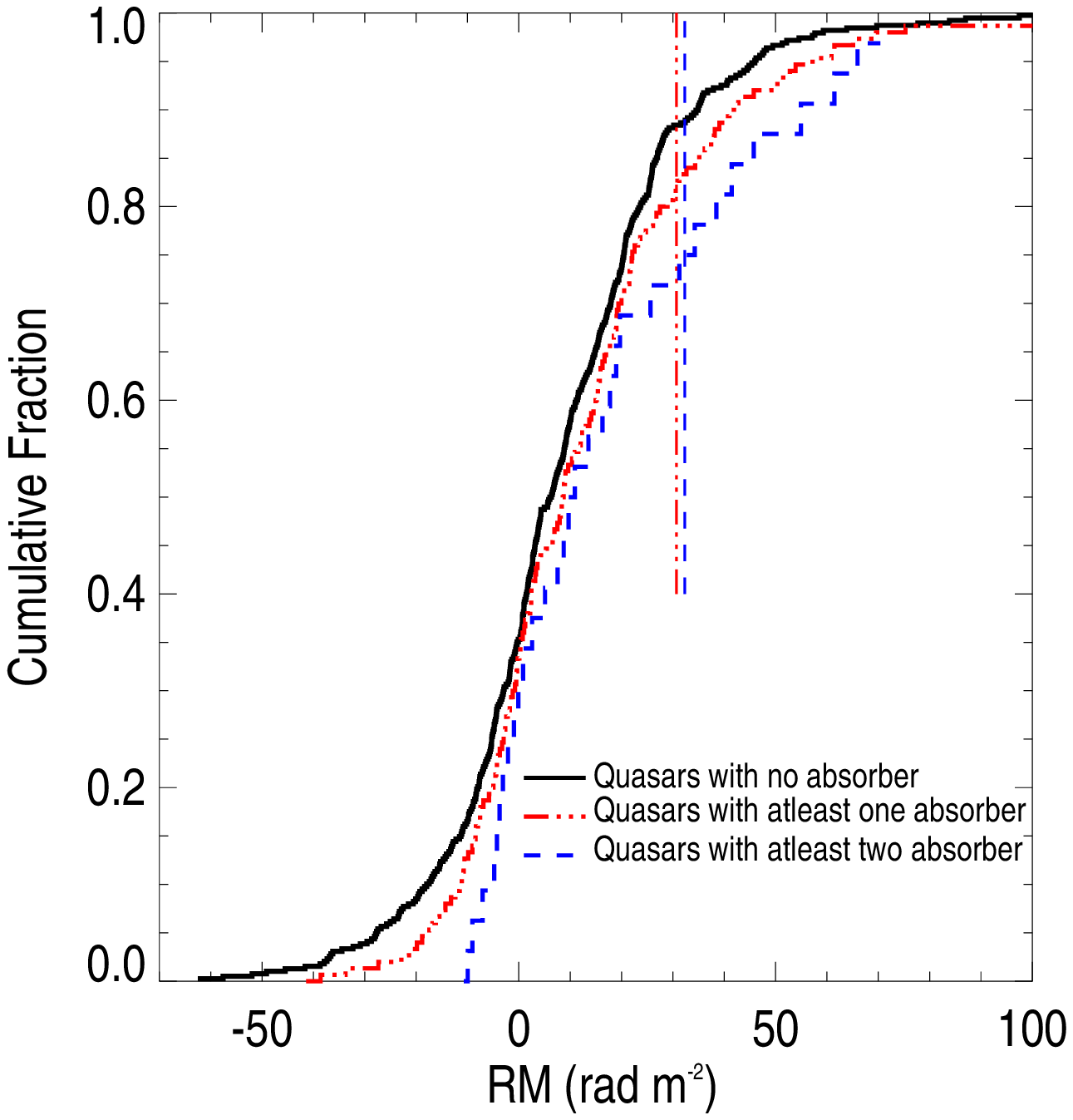,height=8.0cm,width=9.2cm,angle=0}
 \epsfig{figure=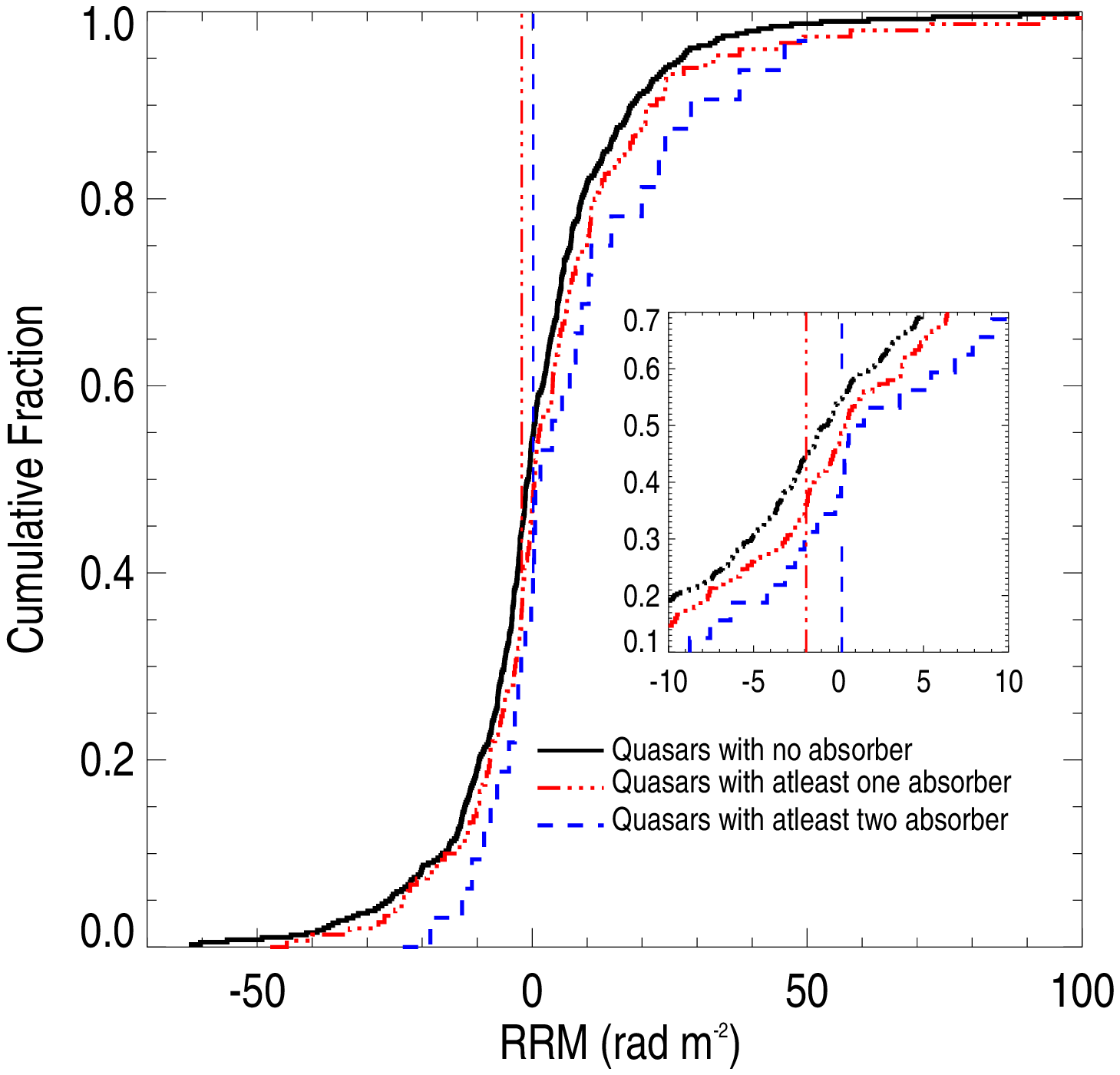,height=7.6cm,width=8.0cm,angle=0}
  \caption{ \emph{Left panel:} Cumulative distribution of the 
    Rotation Measure (RM), for the quasar with no
    absorber (thick line), with at least one absorber (dashed-doted
    line) and with at least two absorber (dashed line). \emph{right
      panel:} same as left, for the Residual
    Rotation Measure (RRM) measurements, the inset displays the
      zoom-in on the maximum distance between the distribution functions. }
 \label{fig:cuml_rrm_zgt1}
 \end{figure*}

\section{Results}
\label{sec:results_frmz}

 The quasars with strong \mgii absorption systems along their
 line-of-sight are found to have broad RM distribution from the
 analysis of 6cm data set~\citep{Bernet2008Natur.454..302B}. However,
 in a recent study no such signature is observed with the 21cm data
 set by \citet[][]{Bernet2012ApJ...761..144B}, though with the nominal
 sample size consisting of only 54 radio source sightlines (see their
 Figure 3). In this work, to test the above discrepancy, in
 Figure~\ref{fig:cuml_rrm_zgt1} (\emph{left panel}), we have shown the
 cumulative distribution of RM for sightlines with \mgii (with
 number $n_{\rm Mg{\sc II}} > 0$, dashed-dotted line; $n_{\rm Mg{\sc
     II}} > 1$, dashed line) and without \mgii ($n_{\rm Mg{\sc II}} =
 0$, thick line) absorber using our large data set of 539 sightlines
 having RM measurement at around 21cm wavelength. The K$-$S test
 rules out the null hypothesis for the quasar subsets, (i) without and
 with at least one \mgii absorber and (ii) without and with at least
 two \mgii absorbers, at a confidence level of 48\% and 66\%
   respectively, which is statistically nonsignificant using our
   this modest sample size. We also notice here that the major difference in
 RM distribution can be seen at around 30~\rrm, which is similar
 to the average contribution of 20~\rrm\ from the foreground galactic
 RM (GRM) component~\citep{Bernet2008Natur.454..302B}. To see any such
 effect of GRM, we have plotted the cumulative distribution of RRM
 (i.e., GRM subtracted RM), shown in the \emph{right panel} of
 Figure~\ref{fig:cuml_rrm_zgt1}. The two distribution without and with
 at least one \mgii absorber are found to be different with the K$-$S
 test giving $P_{null}=0.21$. Similarly, a $P_{null}=0.32$, is seen
 between the quasar subsets having at least two \mgii absorber and no
 absorber, implying that the hypothesis of quasars with and without
 \mgii absorber have a similar distribution is ruled out at a
 confidence level of 79\%.

\begin{figure}
 \epsfig{figure=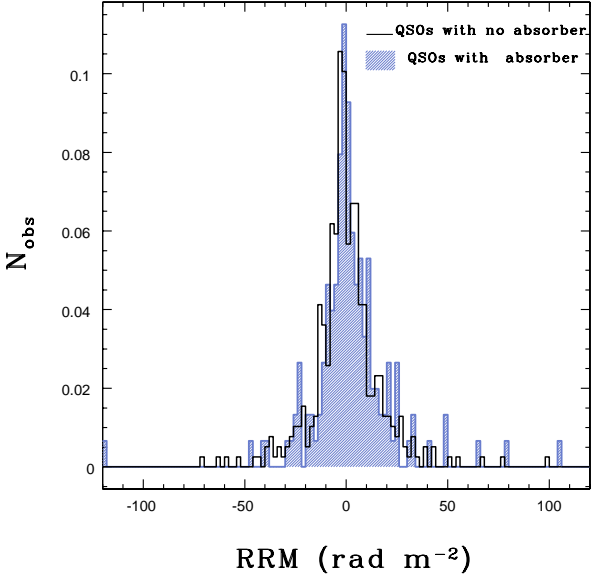,height=8.5cm,width=8.5cm,angle=0}
  \caption{ Histogram of Residual Rotation Measure
    (RRM), normalized by the total quasars counts in the subsets of
    quasar with (\ewr(2796) $\ge$ 0.3\AA) \mgii absorber (shaded
    region) and without absorber (thick line).}
 \label{fig:hist_rrm}
 \end{figure}

To elucidate the effect of intervening absorber on RRM, in
Figure~\ref{fig:hist_rrm}, we have plotted the histogram of RRM for
the quasars with absorber (shaded region) and without absorber (thick
line), after normalizing with total quasar count within respective
subsets. At first look, it appears that the RRM distribution for the
sightlines that passes through the \mgii systems is significantly
broader than that for which absorption is absent. This
reflects in the standard deviation of RRM for the quasars with and
without \mgii absorber being 18.93$\pm1.05$~\rrm\ and
  17.11$\pm0.69$~\rrm, respectively.  Here, the error on standard
  deviation of RRM is computed by propagation of errors in individual
  RRM value by assuming their Gaussian nature, as follows : 

\par


\begin{equation}
\delta \sigma = \frac{1}{\sigma \rm {(N-1)}} \sqrt { \sum_{{\rm
      i}={\rm 1}}^{\rm N}
  {\left( x_i - \overline{x}\right) ^2} \delta x_i^2  + 
\left( \sum_{{\rm i}={\rm 1}}^{\rm N}  x_i - \overline{x}\right) ^2
\delta \overline {x}^2  } ,  
\label{eq:std_err_pro}
\end{equation}

 Where, $\sigma$ is the standard deviation of RRM,
  ${x}_i$'s are the observed RRM values, $\overline{x}$ is mean RRM
value and $\delta{x}_i$'s, $\delta \overline{x}$ are their respective
  errors.

 \begin{figure}
\epsfig{figure=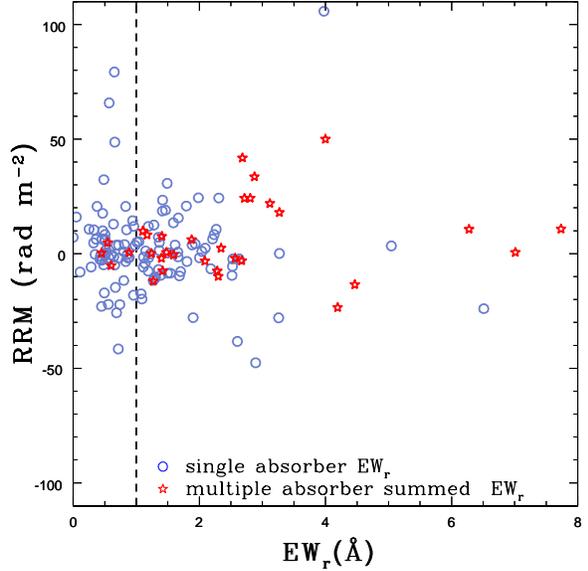,height=8.cm,width=8.cm,angle=0}
  \caption{Distribution of rest frame equivalent width with residual rotation
    measure (RRM) for the quasars with \mgii absorber.}
 \label{fig:rrm_vs_ew}
 \end{figure}

 \begin{figure}
\epsfig{figure=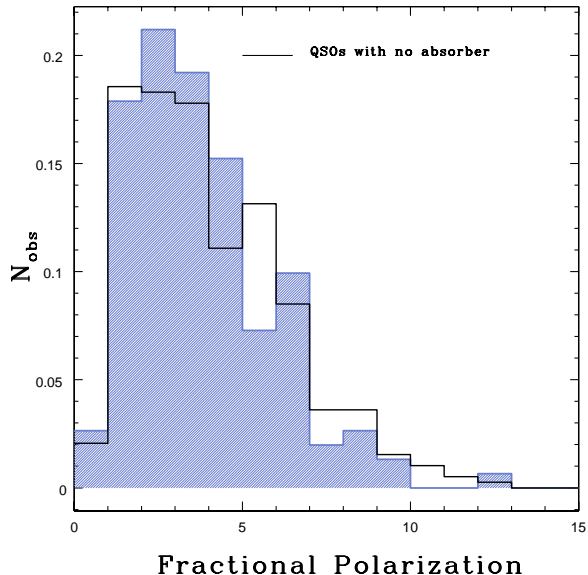,height=8.cm,width=8.cm,angle=0}
  \caption{Histogram of the fractional polarization ($p$) for the quasars
    with \mgii absorber (shaded region) and without absorber
    (thick line).}
 \label{fig:pol_frac}
 \end{figure}

 To quantify the excess standard deviation ($\sigma^{ex}$)
  seen along the sightlines with intervening absorbers, we have
  subtracted the standard deviation of RRM for the quasar subsample with
  and without \mgii absorber in the quadrature i.e.
  
\begin{equation}
 \sigma^{ex} =  \sqrt {\sigma_{{(\rm MgII)}}^2 -  \sigma_{(\rm noMgII)}^2 } ,  
\label{eq:std_excess}
\end{equation}

 The excess in the standard deviation is found to be
8.11$\pm2.85$~\rrm, i.e., at $2.8\sigma$ level, where the associated
error is computed with error propagation as

\begin{equation}
\delta \sigma^{ex} = \frac{1}{\sigma^{ex}} \sqrt {
  \sigma_{{(\rm MgII)}}^2 \delta \sigma_{(\rm MgII)}^2 +
  \sigma_{(\rm noMgII)}^2 \delta   \sigma_{(\rm noMgII)}^2 } ,  
\label{eq:std_excess_pro}
\end{equation}

  This high significance excess at $2.8\sigma$ level in standard
  deviation is much higher than the above discussed $79$\% confidence
  level implied by the K$-$S test for the subsample of quasar with and
  without \mgii absorber. One possibility could be that the error bars
  in RRM measurements may be underestimated. To quantify it, we have
  computed the reduced \chisq~value for our subsample without \mgii
  absorber, by assuming the expected value of RRM measurements to be
  (i) the mean value of RRM measurements, and (ii) as zero value. In
  both the cases, the reduced \chisq~values are found to be around
  2.87. Now, assuming that for true error bars in RRM measurements for
  subsample without \mgii absorber should result this reduced
  \chisq~to be around unity, suggests us to scale up all the error
  bars on our RRM measurements by square root of 2.87. Repeating the
  above analysis with these scaled error bars on RRM, our revised
  standard deviation for quasars with and without \mgii absorber
  becomes 18.93$\pm$1.78~\rrm~and 17.11$\pm$1.16~\rrm~respectively.
  Similarly, the excess standard deviation (using
  Eq~\ref{eq:std_excess}) now become 8.11$\pm$4.83~\rrm, i.e., at
  1.7$\sigma$ level. Given that our K-S test also give the similar
  confidence level (i.e., 79\%) it appear that our earlier result with
  higher significance of $2.8\sigma$ level perhaps may be due to this
  underestimation of error bars in RRM measurements. Therefore,
  henceforth throughout the analysis, we have derived our results
  using the scaled error bars in RRM measurements.

  We also note that our result is in good agreement with
  ~\citet{Hammond2012arXiv1209.1438H} and
  ~\citet{Schnitzeler2010MNRAS.409L..99S}, where they have shown the
  standard deviation of RRM by its extragalactic component in the
  order of 10$-$15 \rrm\ and $\sim$6~\rrm, respectively. In this
  10$-$15 \rrm\ standard deviation of RRM by
  ~\citet{Hammond2012arXiv1209.1438H}, they have also taken into account
  the possible contribution caused by errors associated with GRM and
  RM measurements, by their quadrature subtraction from the observed
  $\sigma$(RRM) value (e.g., see Section~\ref{sec:intro_frm}). Recall
  that in their study they have used a mixture of sightlines
  consisting of both with and without \mgii absorber. However in our
  study, as we do have separate subsample of sightlines with and without
  \mgii absorber, it allows us to quantify the contribution in
  $\sigma^{ex}$ due to intervening absorbers by subtracting the
  $\sigma$(RRM) of later from the former in quadrature (e.g see Eq.
  \ref{eq:std_excess}). This being the difference of two standard
  deviations in quadrature, makes our result of 8.11$\pm4.83$~\rrm,
  free from any error contribution to $\sigma$(RRM) that are common
  for both the subsamples of with and without \mgii absorbers, such as
  associated with the GRM and RM measurements considered in
  ~\citet{Hammond2012arXiv1209.1438H}.

\par

Now, it is also worth to see the RRM distribution with \ewr, as the
quasars having absorber with larger \ewr\ are expected to have larger
observed RRM dispersion. In Figure~\ref{fig:rrm_vs_ew}, we have
plotted the RRM distribution with \ewr\ where the open circle
represents the \ewr\ corresponding to an absorber over a sightline,
while the star (red) corresponds to the sum of \ewr\ values for the
sightlines having more than one \mgii absorber. In the absence of
obvious trend of RRM with \ewr, we have computed the standard
deviation in RRM for the absorber having \ewr\ $<$ 1\AA\ and
\ewr\ $\ge$ 1\AA, which are 18.65$\pm2.42$ and
  19.23$\pm3.23$~\rrm, respectively, with error bars computed using
  Eq~\ref{eq:std_err_pro}. Though there is a very mild excess in the
value of standard deviation of RRM for larger \ewr\ subsample but it
is consistent within $1\sigma$ error bars. \par

 ~\citet{Hammond2012arXiv1209.1438H} found an anti-correlation between
RRM and fractional polarization ($p$). One possible physical bias
suggested by them was the presence of intervening absorbers; which
need to be tested by plotting the RRM with `$p$' for a subclass of
sample with and without \mgii absorber. In this possible scenario, to
explain anticorrelation of RRM versus $p$, one would expect that the
fractional polarization observed for a sample with \mgii absorber
(having larger RRM) should be smaller than the sample without \mgii
absorber. To test this hypothesis in Figure~\ref{fig:pol_frac}, we
have plotted the histogram of fractional polarization ($p$) for the
quasars with (shaded region) and without (thick line) \mgii absorber.
The median (weighted mean) values of $p$, for sample with and without
\mgii absorber, are found to be 3.50(4.57) and 3.57(5.16),
  respectively. The K$-$S test shows that the null hypothesis is
ruled out with a low confidence level of 63.13\%. This very
  mild difference with present sample size does not allow us to
  conclude much on above hypothesis, but a larger sample 
  will be helpful to say firmly about it.

\section{Discussion and Conclusions}
\label{sec:discussion}

The excess extragalactic  FR due to the presence of a strong
intervening absorber in the optical spectrum of quasars is noticed in
many recent studies~\citep{Kronberg1982ApJ...263..518K,
  Welter1984ApJ...279...19W,
  You2003AcASn..44S.155Y,Bernet2008Natur.454..302B}. Commonly accepted
view is that the major contributor could be the intervening galaxies
along the sightline of polarized radio sources. For instance,
\citet[][]{Bernet2008Natur.454..302B}  have used a sample of total 76
sightlines with RM observations at around 6cm, and found a
statistically higher RM for their 7 sources showing more than one
\mgii absorber. Many other studies have also shown the increase of RRM
with
redshift~\citep[e.g.,][]{Kronberg2008ApJ...676...70K,Kronberg1976Natur.263..653K,
  Rees1972A&A....19..189R}, albeit by using only a nominal sample size
till now. Recently, with a large data set of 3651 quasars with RRM
observation at 21cm,~\citet{Hammond2012arXiv1209.1438H} have not found
any such signature of RRM evolution with redshift. Similarly, the role
of intervening absorber in RRM at 21cm was also studied till now by
using a small sample size of only 54
quasars~\citep{Bernet2012ApJ...761..144B}. Here, we have addressed
these questions by using a larger sample size of the 539 SDSS quasars
having RM data set at wavelength at around 21cm, rather than at around
6cm .

 Our analysis has shown that RRM distribution for the quasars with and
 without \mgii absorber do differ at confidence level of 79\%. This
 show that there is contribution of intervening absorber to enhance
 the RRM, even at wavelength of 21cm. Its non-detection in the
 analysis of~\citet{Bernet2012ApJ...761..144B}, could be due to their
 smaller sample size, where they have used just a sample of 54
 sightlines having RRM measurement at 21cm. However, we also note that
 at 6cm measurement of RM, the two distribution with and without
 \mgii absorber do differ with much more confidence levels
 ~\citep[e.g,][]{Bernet2008Natur.454..302B} than what we have found
  for  our sample of RRM measurement at 21cm. This may be due to the
 recent mechanism proposed by~\citet{Bernet2012ApJ...761..144B}, where
 the intervening absorbers acts as a RM screens and causes more
 depolarization at longer wavelength. \par

We also found that the standard deviation of RRM for sample with \mgii
absorber do have excess of about 8.11$\pm4.83$~\rrm\ as compared to
the sample without \mgii absorber. This is consistent
with~\citet{Hammond2012arXiv1209.1438H} and
~\citet{Schnitzeler2010MNRAS.409L..99S} studies, where in their
independent analysis they have found the value of extragalactic
component in RRM standard deviation to be around 10$-$15~\rrm\ and
6~\rrm\ respectively.

 Further, we have also computed the standard deviation of RRM for
 subsets having \mgii absorber with \ewr\ $<$ 1\AA\ and \ewr\ $\ge$
 1\AA, as another check to the hypothesis of contribution by
 intervening absorber, where for stronger absorber RRM dispersion is
 expected to be higher than weaker absorber. The values for \ewr\ $<$
 1\AA\ and \ewr\ $\ge$ 1\AA\ are found to be 18.65$\pm2.42$ and
   19.23$\pm3.23$~\rrm, respectively (see
 Figure~\ref{fig:rrm_vs_ew}), which is tentatively consistent with
 above expectation.

We also notice that the factional polarization ($p$) of the quasars
with and without \mgii absorber show nominal difference, with median
(weighted mean) values of $p$, to be 3.50(4.57) and 3.57(5.16)
respectively. This very nominal excess for sightlines without \mgii
absorbers, could be due to the less depolarization in the absence of
intervening absorbers~\citep[e.g., see
  also][]{Hammond2012arXiv1209.1438H,Bernet2012ApJ...761..144B}.
However, a larger data set would require to reduce the statistical
error bar.\par

At present, with our large data set of 539 sightlines, we could only
found a statistical excess of 8.11$\pm4.83$~\rrm\ in standard
deviation of RRM for the sample with \mgii absorber as compared to the
sample without \mgii absorber, by using RRM observation at around
  21cm wavelength. This has allowed us to conclude that the
intervening \mgii absorber also makes contribution for the excess in
standard deviation of RRM observed even at around 21cm wavelength,
though perhaps with smaller magnitude than at
6cm~\citep[e.g,][]{Bernet2008Natur.454..302B}; the theoretical
implication of which needs further explorations.

\section*{Acknowledgments}
 We thank the anonymous referee for the constructive 
  and helpful suggestions. We also thank R. Srianand for useful
discussion. \par
 Funding for the SDSS and SDSS-II has been provided by the Alfred P.
 Sloan Foundation, the Participating Institutions, the National
 Science Foundation, the U.S. Department of Energy, the National
 Aeronautics and Space Administration, the Japanese Monbukagakusho,
 the Max Planck Society, and the Higher Education Funding Council for
 England. The SDSS Web Site is http://www.sdss.org/. The SDSS is
 managed by the Astrophysical Research Consortium for the
 Participating Institutions. The Participating Institutions are the
 American Museum of Natural History, Astrophysical Institute Potsdam,
 University of Basel, University of Cambridge, Case Western Reserve
 University, University of Chicago, Drexel University, Fermilab, the
 Institute for Advanced Study, the Japan Participation Group, Johns
 Hopkins University, the Joint Institute for Nuclear Astrophysics, the
 Kavli Institute for Particle Astrophysics and Cosmology, the Korean
 Scientist Group, the Chinese Academy of Sciences (LAMOST), Los Alamos
 National Laboratory, the Max-Planck-Institute for Astronomy (MPIA),
 the Max-Planck-Institute for Astrophysics (MPA), New Mexico State
 University, Ohio State University, University of Pittsburgh,
 University of Portsmouth, Princeton University, the United States
 Naval Observatory, and the University of Washington.

\label{lastpage}

\bibliography{references}
\end{document}